\documentclass[letterpaper, 10 pt, conference]{ieeeconf}  % Comment this line out if you need a4paper

\IEEEoverridecommandlockouts                              % This command is only needed if 
\usepackage{graphicx, caption, subcaption}      % include this line if your document contains figures
\usepackage{nicefrac}
\let\labelindent\relax
\usepackage{enumitem}
\usepackage{amsmath,amssymb,amsfonts}
\usepackage{mathtools}
\usepackage{physics}
\usepackage{pgfplots}
\usepackage{hyperref}
\usepackage{siunitx}
\pgfplotsset{
	layers/my layer set/.define layer set={
		background,
		main,
		foreground
	}{
	},
	set layers=my layer set,
}
\usepackage{multirow}
\usepackage{caption}
\usepackage{siunitx} 
\usepackage{array}
\usepackage[nolist]{acronym}
\usepackage{mathptmx} % assumes new font selection scheme installed
\usepackage{times} % assumes new font selection scheme installed

\usepackage[backend=bibtex,style=trad-plain]{biblatex}
\title{\LARGE \bf
Adaptive Model Predictive Control for Differential-Algebraic Systems towards a Higher Path Accuracy for Physically Coupled Robots
}

\author{Xin Ye$^{1}$, Karl Handwerker$^{1}$ and S{\"o}ren Hohmann$^{2}$% <-this % stops a space
\thanks{$^{1}$Xin Ye and Karl Handwerker are with FZI Research Center for Information Technology,	76131 Karlsruhe, Germany	
        {\tt\small ye, handwerker@fzi.de}}%
\thanks{$^{2}$S{\"o}ren Hohmann is with Institute of Control Systems, Karlsruhe Institute of Technology, 76131 Karlsruhe, Germany
        {\tt\small soeren.hohmann@kit.edu}}%
}

\begin{document}
\begin{acronym}
	\acro{DOF}{degree of freedom}
	\acro{DAE}{differential algebraic equations}
	\acro{MPC}{model predictive controller}
	\acro{TCP}{tool center point}
	\acro{RBD}{rigid-body dynamics}
\end{acronym}

\maketitle
\thispagestyle{empty}
\pagestyle{empty}

%%%%%%%%%%%%%%%%%%%%%%%%%%%%%%%%%%%%%%%%%%%%%%%%%%%%%%%%%%%%%%%%%%%%%%%%%%%%%%%%
\begin{abstract}

The physical coupling between robots has the potential to improve the capabilities of multi-robot systems in challenging manufacturing processes. 
However, the path tracking accuracy of physically coupled robots is not studied adequately, especially considering the uncertain kinematic parameters, the mechanical elasticity, and the built-in controllers of off-the-shelf robots.
This paper addresses these issues with a novel differential-algebraic system model which is verified against measurement data from real execution. 
The uncertain kinematic parameters are estimated online to adapt the model.
Consequently, an adaptive model predictive controller is designed as a coordinator between the robots. 
The controller achieves a path tracking error reduction of 88.6\% compared to the state-of-the-art benchmark in the simulation.

\end{abstract}

\section{Introduction}
\begin{figure}[t]%{.49\textwidth}%
	\begin{center}
		\includegraphics[width=8cm]{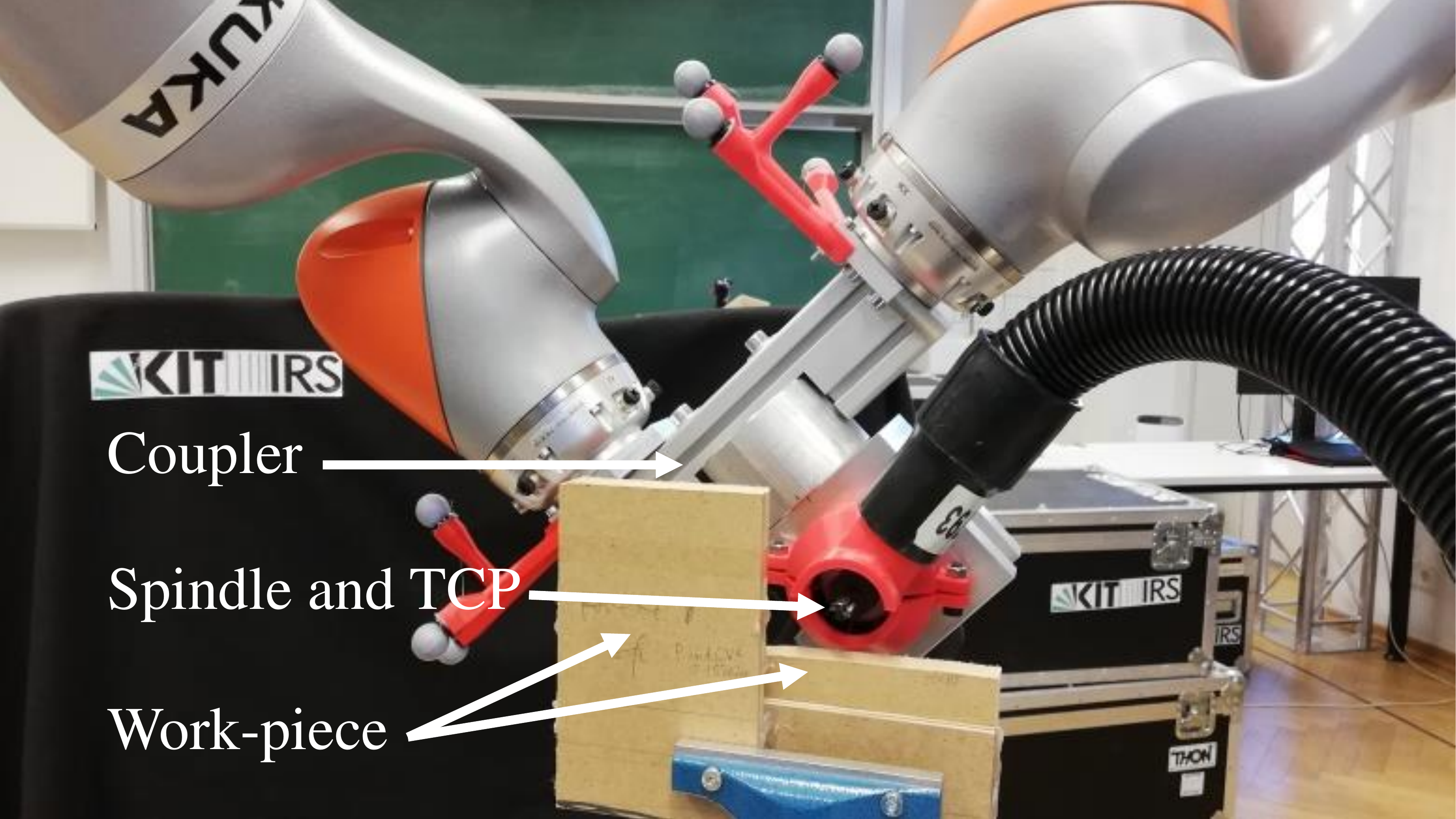}
		\captionof{figure}{Two robots are coupled to a coupler. The spindle and \ac{TCP} are mounted on the coupler to achieve a high stiffness while cutting the work-piece.}
		\label{fig:coupled_robots}
		\vspace*{-5mm}
	\end{center}
\end{figure}

Robot manipulators are becoming increasingly important in manufacturing processes.
Apart from the conventional tasks, such as object handling, welding, or surface painting, it has been attempted to utilize robots in more challenging processes such as milling and sheet metal bending \cite{abele2007modeling}. 
The goal is to replace the expensive Computerized Numerical Control (CNC) machines with robots by exploiting their versatility and low cost \cite{perzylo2019smerobotics}.
Due to the limitation of robots in stiffness and load capability under process forces, the idea of physically coupling multiple robots to a single end-effector (also named as coupler below) is proposed \cite{muhlbeier2020value}, as illustrated in Fig.~\ref{fig:coupled_robots}. 
This idea inherits the research work in the cooperative manipulation \cite{caccavale2016cooperative} since the 1970s, but brings about new aspects in stiffness \cite{ye2022stiffness} and accuracy \cite{ye2023enhancement} for physically coupled robots which are essential for manufacturing processes.

Our previous work \cite{ye2023enhancement} analyses the cause of inaccuracy in the unknown kinematic parameters of robot base placements and the coupler geometry, as well as in the imperfect synchronization of joint trajectories. 
These factors lead to internal stress between the coupled robots, deformation of robots' joints, and consequently the deviation of the \ac{TCP} from the path. 
As the tracking accuracy along a pre-defined path is improved by an offline compensation method using the measured data from a non-compensated trial execution in \cite{ye2023enhancement}, neither the offline compensation nor the trial execution is needed in the present work.
Instead, the accuracy enhancement is brought to a potentially online-capable control framework.
In order to make our method applicable to the manufacturing practice, the following requirements should be satisfied:
\begin{itemize}
	\item A given path with arbitrary complexity should be tracked by the \ac{TCP} with high accuracy;
	\item Deviations in the joint trajectories should be corrected instantly during the execution;
	\item The uncertain kinematic parameters should be estimated with actual sensor measurements and updated to the utilized model for the adaptive control instantly; 
	\item The method should be applicable to real robots with non-negligible elasticity, particularly in joints \cite{abele2007modeling};
	\item The built-in controller of off-the-shelf robots should not be circumvented;
	\item Additional objectives such as the load distribution and the limitation of internal stress between robots should be handled optimally.
\end{itemize}

In the literature of control regarding cooperative manipulation, the above requirements cannot be satisfied simultaneously, because the accuracy of physically coupled robots is not studied adequately. 
Preliminary works \cite{aghili2005unified,meng2022passivity} solve the position and internal force control problem for constrained multibody systems, where passive joints or joints with elasticity are considered. 
These methods have not been applied to real robots because the control is not adaptable to uncertain kinematic parameters, upon which the internal force and joint loads are highly sensitive.
Adaptive control is applied in \cite{aghili2012adaptive,erhart2016cooperative} for uncertain kinematic parameters to achieve a minimal internal stress or positioning accuracy. 
However, these methods have difficulty on real robots because the limited joint stiffness cannot be considered in the controllers.
Black- or grey-box adaptive control methods based on neural networks \cite{panwar2012adaptive,xu2018adaptive} or inverse dynamics learning \cite{reuss2022end} have the potential to approximate the robots' behavior in physical coupling, but the high computational load and the limited ability of generalization still hinder their utilization.
Furthermore, the aforementioned studies assume a direct access to the torque input of joints, which is only available for robot prototypes instead of off-the-shelf robots.

In view of the limitation in state-of-the-art methods, the contributions of the present work are as follows:
\begin{itemize}
	\item A novel model of \ac{DAE} for the physically coupled robots which accounts for the joint stiffness and the uncertain kinematic parameters
	\item The reduction of the differentiation order of the \ac{DAE} model by considering the dynamics of the built-in impedance controller of off-the-shelf robots
	\item The estimation of uncertain kinematic parameters using real measurement data based on the sensitivity analysis of the \ac{DAE} 
	\item  An adaptive \ac{DAE} \ac{MPC} that corrects the joint trajectories instantly towards a higher path tracking accuracy
\end{itemize}
%Section~\ref{sec:model} describes the \ac{DAE} model. 
%\input{tex/02_related_work}
\section{Modeling of Physically Coupled Robots}\label{sec:model}
This section describes the modeling of the physically coupled robots in \ac{DAE} as the basis of the adaptive \ac{MPC}. The model incorporates the \ac{RBD} of each individual robots, the joint elasticity in addition to the \ac{RBD}, the coupling constraints, as well as the characteristics of the built-in controller of an off-the-shelf robot type KUKA LBR iiwa 14 R820, abbreviated as LBR. 

\subsection{Dynamics in the Physical Coupling}\label{subsec:alg}
\begin{figure}[h]%{.49\textwidth}%
	\begin{center}
		\vspace*{3mm}
		\includegraphics[width=8cm]{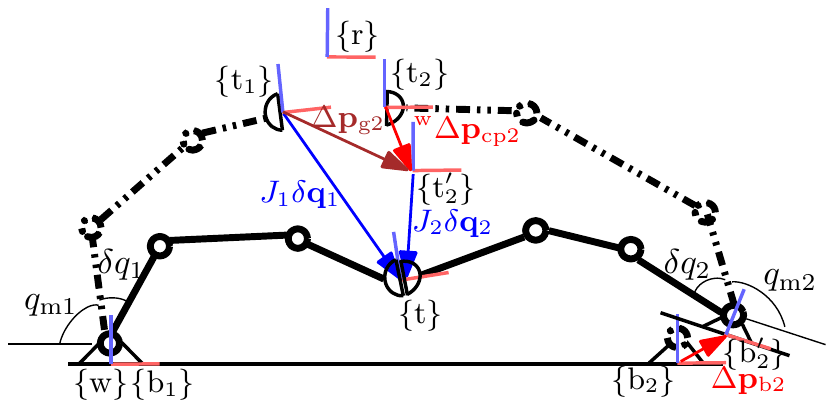}
		\captionof{figure}{The loop closure condition of a two-robot example}
		\label{fig:couple_sketch}
		\vspace*{-5mm}
	\end{center}
\end{figure}

A part of the error analysis from \cite{ye2023enhancement} is recapitulated and slightly modified for the novel \ac{DAE} model. 
Consider a system of $n_\mathrm{R}$ physically coupled robots, each having $n_\mathrm{J}$ joints. 
An example of two robots is shown in Fig.~\ref{fig:couple_sketch}.
The dash-double-dotted lines depict the nominal configurations of robots at the measured motor driving side joint angles $\boldsymbol{q}_{\mathrm{m}i}\in\mathbb{R}^{n_\mathrm{J}},i=1,\dots,n_\mathrm{R}$.
Due to the elasticity and the physical coupling, the configurations at the driven link side, depicted in solid lines, are not identical to the motor driving side. 
The joints of robot~$i=1,\dots,n_\mathrm{R}$ undergo elastic deformation $\delta\boldsymbol{q}_i\in\mathbb{R}^{n_\mathrm{J}}$ that complies to Hooke's Law by:
\begin{align}\label{eq:joint_elas}
	\boldsymbol{\tau}_{\mathrm{m}i}=K_\mathrm{J}\delta\boldsymbol{q}_i,
\end{align}
where the joint torque $\boldsymbol{\tau}_{\mathrm{m}i}\in\mathbb{R}^{n_\mathrm{J}}$ is measured by the torque sensors.
$K_\mathrm{J}$ is a diagonal matrix with constant joint-stiffness parameters in diagonal entries.

%\begin{table}[h]
%	\caption{Frame notation}
%	\begin{tabular}{lllllllll} 
%		Frame & $\{\mathrm{w}\}$ & $\{\mathrm{b}_i\}$&  $\{\mathrm{t}\}$&$\{\mathrm{r}\}$\\ \hline
%		Description & world & base & TCP&reference
%	\end{tabular}\label{tab:notations}
%\end{table}
The coupling compels the robots to share a common \ac{TCP} frame \{t\}, which is deviated away from the reference path point \{r\}.
To enhance precision, it is crucial to identify and immediately mitigate the factors contributing to this deviation. To achieve that, the source of errors, namely the uncertain kinematic parameters in the coupling, is analysed as follows.
In the world frame \{w\}, we now observe robot $i=2,\dots,n_\mathrm{R}$, which corresponds to the one on the right-hand-side of Fig.~\ref{fig:couple_sketch}. 
From the nominal placement frame $\{\mathrm{b}_i\}$, the real base $\{\mathrm{b}_i'\}$ is deviated by a small error $\Delta\boldsymbol{p}_{\mathrm{b}i}\in\mathbb{R}^6$ in rotational and translational dimensions.
From $\{\mathrm{b}_i'\}$ to the robots' nominal \ac{TCP} frame $\{\mathrm{t}_i\}$, the transformation is determined by the forward kinematics $T_{\mathrm{kin}i}(\boldsymbol{q}_{\mathrm{m}i})$. 
The geometric error of the coupler $~^\mathrm{w}\Delta\boldsymbol{p}_{\mathrm{cp}i}\in\mathbb{R}^6$ (after its transformation from the coupler frame constant vector $~^{\mathrm{body}}\Delta\boldsymbol{p}_{\mathrm{cp}i}$ to the world frame) is superposed upon $\{\mathrm{t}_i\}$, resulting in $\{\mathrm{t}_i'\}$. 
If the uncertain kinematic parameters are summarized in the vector $\boldsymbol{p}_i=[\Delta\boldsymbol{p}_{\mathrm{b}i}^T,~^{\mathrm{body}}\Delta\boldsymbol{p}_{\mathrm{cp}i}^T]^T$, then the transformation from \{w\} to $\{\mathrm{t}_i'\}$ is denoted as:
\begin{align}
	~^\mathrm{w}T_{\mathrm{t}_i'}(\boldsymbol{q}_{\mathrm{m}i},\boldsymbol{p}_i)=~^\mathrm{w}T_{\mathrm{b}_i}~^{\mathrm{b}_i}T_{\mathrm{b}_i'}(\Delta\boldsymbol{p}_{\mathrm{b}i})T_{\mathrm{kin}i}(\boldsymbol{q}_{\mathrm{m}i})~^{\mathrm{t}_i}T_{\mathrm{t}_i'}(^{\mathrm{body}}\Delta\boldsymbol{p}_{\mathrm{cp}i})
\end{align}

W.l.o.g., the base of the first robot $\{\mathrm{b}_1\}$ is defined on the origin of world frame $\{\mathrm{w}\}$ without placement error. 
Also as in \cite{ye2023enhancement}, the first robot is the only one that is assumed to be well calibrated w.r.t. the \ac{TCP}.
Therefore, it applies $~^\mathrm{w}T_{\mathrm{t}1}(\boldsymbol{q}_{\mathrm{m}1})=T_{\mathrm{kin}1}(\boldsymbol{q}_{\mathrm{m}1})$.
As such, the gap vector $\Delta \boldsymbol{p}_{\mathrm{g}i}(\boldsymbol{q}_{\mathrm{m}1},\boldsymbol{q}_{\mathrm{m}i},\boldsymbol{p}_i)\in\mathbb{R}^6,\forall i=2,\dots,n_\mathrm{R}$ can be computed as the difference between $~^\mathrm{w}T_{\mathrm{t}_i'}(\boldsymbol{q}_{\mathrm{m}i},\boldsymbol{p}_i)$ and $~^\mathrm{w}T_{\mathrm{t}1}(\boldsymbol{q}_{\mathrm{m}1})$, denoted as $\Delta \boldsymbol{p}_{\mathrm{g}i}=\mathrm{diff}(~^\mathrm{w}T_{\mathrm{t}_i'},~^\mathrm{w}T_{\mathrm{t}1})$ where the operator $\mathrm{diff}(\cdot,\cdot)$ is define as follows:
\begin{align}\label{eq:twist_difference}
	\mathrm{diff}(T_1,T_2)=\begin{bmatrix}\frac{1}{2}\left(R_2^T R_1-R_1^T R_2\right)^{\lor}\\
		\boldsymbol{t}_1-\boldsymbol{t}_2\end{bmatrix}
	\end{align}
Whereby, the rotation matrices $R$ and the translation vectors $\boldsymbol{t}$ are extracted from the corresponding matrix $
T=\begin{bmatrix}R&\boldsymbol{t}\\\boldsymbol{0}^T&1\end{bmatrix}\label{eq:hom_rot_tr}$.
According to \cite{bullo2005geometric,lee2010geometric}, the \textit{vee map} $\lor:so(3)\rightarrow\mathbb{R}^3$ is the inverse of the hat map for a skew-symmetric matrix.
This convention is adopted to facilitate a symbolic computation in solving the subsequent estimation and control problem.

%$\Delta\boldsymbol{p}=[\Delta\boldsymbol{p}_{\mathrm{b}2}^T,\dots,\Delta\boldsymbol{p}_{\mathrm{b}n_\mathrm{R}}^T,\Delta\boldsymbol{p}_{\mathrm{cp}2}^T,\dots,\Delta\boldsymbol{p}_{\mathrm{cp}n_\mathrm{R}}^T]^T$
%=\begin{bmatrix}\Delta \alpha_{\mathrm{t}_i}\\\Delta\gamma_{\mathrm{t}_i}\\\Delta \beta_{\mathrm{t}_i}\\\Delta x_{\mathrm{t}_i}\\\Delta y_{\mathrm{t}_i}\\\Delta z_{\mathrm{t}_i}\end{bmatrix}&

As depicted in Fig.~\ref{fig:couple_sketch}, the coupling condition can finally be described by the loop closure equation:
\begin{align}\label{eq:loop_closure}
	J_1\delta\boldsymbol{q}_1=\Delta \boldsymbol{p}_{\mathrm{g}i}(\boldsymbol{q}_{\mathrm{m}1},\boldsymbol{q}_{\mathrm{m}i},\boldsymbol{p}_{i})+J_i\delta\boldsymbol{q}_i,i=2,\dots,n_\mathrm{R}
\end{align}
where $J_i$ is the Jacobian matrix.

While the elasticity between the driving motor side and the driven link side of each joint is considered, the links are assumed rigid. 
Therefore, \ac{RBD} still applies to the driven link side of the joints for each individual robot $i=1,2,\dots,n_\mathrm{R}$:
\begin{align}
	M_i\ddot{\boldsymbol{q}}_i+\boldsymbol{h}_i+\boldsymbol{g}_i=J_i^T\boldsymbol{\lambda}_i+ \boldsymbol{\tau}_{i}+\boldsymbol{e}_{\mathrm{d}i}\label{eq:dyn_r1}
\end{align}
Whereby, $\boldsymbol{q}_i\in\mathbb{R}^{n_\mathrm{J}}$ is the link side joint angle obtained from $\boldsymbol{q}_i=\boldsymbol{q}_{\mathrm{m}i}+\delta\boldsymbol{q}_{i}$, $M_i$ is the joint space inertia matrix, $\boldsymbol{h}_i\in\mathbb{R}^{n_\mathrm{J}}$ includes Coriolis and centrifugal torques, $\boldsymbol{g}_i\in\mathbb{R}^{n_\mathrm{J}}$ is the gravity torque. 
$\boldsymbol{\lambda}_i\in\mathbb{R}^6$ is the constraint wrench between the coupler and robot $i$.
$\boldsymbol{\tau}_{i}\in\mathbb{R}^{n_\mathrm{J}}$ is the torque vector provided by the motors and the corresponding reduction gears in joints, whereas torque sensors measure the vector $\boldsymbol{\tau}_{\mathrm{m}i}=-\boldsymbol{\tau}_{i}$.
Frictions and other errors in the dynamics are combined in $\boldsymbol{e}_{\mathrm{d}i}\in\mathbb{R}^{n_\mathrm{J}}$.
By combining the first three terms of (\ref{eq:dyn_r1}) into $\boldsymbol{\tau}_{\mathrm{RBD}i}\in\mathbb{R}^{n_\mathrm{J}}$ and neglecting $\boldsymbol{e}_{\mathrm{d}i}$, one obtains:
\begin{align}
	\boldsymbol{\tau}_{\mathrm{RBD}i}+\boldsymbol{\tau}_{\mathrm{m}i}=J_i^T\boldsymbol{\lambda}_i\label{eq:dyn_short}
\end{align}

The \ac{RBD} of the coupler is described as follows:
\begin{align}
	\boldsymbol{w}_\mathrm{cp}=\boldsymbol{w}_\mathrm{ext}-\sum_{i=1}^{n_\mathrm{R}}\boldsymbol{\lambda}_i\label{eq:dyn_cp}
\end{align}
Whereby, $\boldsymbol{w}_\mathrm{ext}\in\mathbb{R}^6$ is the external process wrench exerted on \ac{TCP}. 
$\boldsymbol{w}_\mathrm{cp}\in\mathbb{R}^6$ is the combination of inertial and gravity wrench of the coupler.

Combining (\ref{eq:joint_elas}) (\ref{eq:loop_closure}) (\ref{eq:dyn_short}) and (\ref{eq:dyn_cp}), one obtains the algebraic equations in matrix form:
\begin{align}\label{eq:alg}
	F\boldsymbol{x}_\mathrm{a}=\boldsymbol{b}(\boldsymbol{p},\boldsymbol{w}_\mathrm{ext}),
\end{align}
where the algebraic variables are $\boldsymbol{x}_\mathrm{a}=[\boldsymbol{\tau}_\mathrm{m1}^T,\dots,\boldsymbol{\tau}_{\mathrm{m}n_\mathrm{R}}^T,\boldsymbol{\lambda}_1^T,\dots,\boldsymbol{\lambda}_{n_\mathrm{R}}^T]^T$. On the right hand side of (\ref{eq:alg}), $\boldsymbol{b}(\boldsymbol{p},\boldsymbol{w}_\mathrm{ext})=[\boldsymbol{\tau}_\mathrm{RBD1}^T\dots,\boldsymbol{\tau}_{\mathrm{RBD}n_\mathrm{R}}^T,\boldsymbol{w}_\mathrm{cp}^T-\boldsymbol{w}_\mathrm{ext}^T,\Delta \boldsymbol{p}_{\mathrm{g}2}(\boldsymbol{p}_2)^T,\dots, \Delta\boldsymbol{p}_{\mathrm{g}n_\mathrm{R}}(\boldsymbol{p}_{n_\mathrm{R}})^T]^T$. 
$\boldsymbol{p}$ is a concatenation of $\boldsymbol{p}_i,i=2,\dots,n_\mathrm{R}$.
For matrix $F$ in (\ref{eq:alg}), only the formulation for two robots are given to keep the matrix tidy:
\begin{align}\label{eq:F_matrix}
	F=\begin{bmatrix}-I&O&J_1^T&O\\
		O&-I&O&J_2^T\\O&O&-I&-I\\J_1K^{-1}_\mathrm{J}&-J_2K^{-1}_\mathrm{J}&O&O\end{bmatrix}
\end{align}
This matrix is easily extendable to any number of robots.

We do not build the differential equations w.r.t. $\boldsymbol{q}_i$ and its derivatives as in (\ref{eq:dyn_r1}), because $\boldsymbol{q}_i=\boldsymbol{q}_{\mathrm{m}i}+\delta\boldsymbol{q}_{i}=\boldsymbol{q}_{\mathrm{m}i}+K_\mathrm{J}^{-1}\boldsymbol{\tau}_{\mathrm{m}i}$ is readily obtained from sensor measurements and
$\boldsymbol{\tau}_{\mathrm{RBD}i}(\boldsymbol{q}_i,\dot{\boldsymbol{q}}_i,\ddot{\boldsymbol{q}}_i)$ is computed from the inverse dynamics. 
Therefore, $\boldsymbol{\tau}_{\mathrm{RBD}i}$ are regarded as known parameters that are updated in each model prediction loop and placed in the corrsponding entries of $\boldsymbol{b}$. 
The same reason applies to the known parameters $\boldsymbol{w}_\mathrm{cp}$. 
Also $\boldsymbol{w}_\mathrm{ext}$ is available in run time through a conversion from the measurements $\boldsymbol{\tau}_{\mathrm{m}i}$.
% that can be decomposed as $\boldsymbol{w}_\mathrm{cp}=M_\mathrm{cp}\ddot{\boldsymbol{q}}_\mathrm{cp}+\boldsymbol{h}_\mathrm{cp}+\boldsymbol{g}_\mathrm{cp}$.
\subsection{Characterization of the Built-in Controller}
Since the built-in controller of most off-the-shelf robots cannot be circumvented, it is only possible to design a superimposed controller that coordinates all the decentralized built-in controllers of each individual robot. It is therefore necessary to characterize the behaviors of the built-in controller for a complete \ac{DAE} model.

For the studied LBR, three modes are available, namely joint specific position control, joint specific impedance control, and Cartesian impedance control \cite{schreiber2010fast}.
Since the modeling in Section~\ref{subsec:alg} is in the joint space, the Cartesian space control mode is rejected because it brings about complications in the transformation between the spaces.
The joint specific position control exhibits the highest control stiffness, and behaves similar to a proportional–integral–derivative (PID) controller. 
When there is a deviation of joint position from a given set-point and the deviation cannot be eliminated immediately due to the coupling constraint, the integral term of the PID controller accumulates, resulting in a gradual increase in motor torque output over time.
Although this integral term can be seen as a state variable in the differential equation, the built-in controller does not provide its value to the user. 
The joint specific impedance control is chosen as the baseline controller upon which \ac{DAE}-\ac{MPC} is designed, because it lacks the integral term, \ac{RBD} is compensated in its control law, and the stiffness and damping constants of the controller can be set by the user.
The controller stiffness is limited to $\SI{2000}{\newton\meter\per\radian}$ across all joints, rendering it lower than the achievable stiffness offered by the joint specific position control. 
However, a superposed controller is capable to achieve a higher overall controller stiffness, as long as the joint position set-point is moved in the direction against the joint torque, proportional to the effects of the external process wrench $\boldsymbol{w}_\mathrm{ext}$.
%If any controller is able to bring stiffness to an uncontrolled robot with zero stiffness against $\boldsymbol{w}_\mathrm{ext}$, then a controller that is superposed on an already controlled robot must also be able to bring a change to the stiffness.

According to \cite{schreiber2010fast}, the control law of the joint specific impedance control for robot $i=1,\dots,n_\mathrm{R}$ is approximately rewritten with the symbols of the present paper as follows:
\begin{align}\label{eq:imp_ctrl_law}
	\boldsymbol{\tau}_i=K_\mathrm{P}(\boldsymbol{q}_{\mathrm{cmd}i}-\boldsymbol{q}_{\mathrm{m}i})-K_\mathrm{D}\dot{\boldsymbol{q}}_{\mathrm{m}i}+\boldsymbol{\tau}_{\mathrm{FRI}i}+\boldsymbol{\tau}_{\mathrm{RBD,comp,}i}
\end{align}
Whereby, $K_\mathrm{P}\in\mathbb{R}^{n_\mathrm{J}\times n_\mathrm{J}}$ is the diagonal controller stiffness matrix that is set to $\SI{2000}{\newton\meter\per\radian}$ for all joints in the present study.
$\boldsymbol{q}_{\mathrm{cmd}i}\in\mathbb{R}^{n_\mathrm{J}}$ is the joint position command.
The damping term is a linear approximation with the constant diagonal matrix $K_\mathrm{D}\in\mathbb{R}^{n_\mathrm{J}\times n_\mathrm{J}}$.
Although the diagonal entries of $K_\mathrm{D}$ are variable w.r.t. $\dot{\boldsymbol{q}}_{\mathrm{m}i}$, the nominal values are identified as $\SI{10}{\newton\meter\per(\radian\per\second)}$ for all joints at low and moderated magnitude of $\dot{\boldsymbol{q}}_{\mathrm{m}i}$. 
$\boldsymbol{\tau}_{\mathrm{FRI}i}$ is an optional torque overlay given by the user. 
The exploitation of this feature is regarded as a future work to enhance the stiffness of the multi-robot system in milling process, but $\boldsymbol{\tau}_{\mathrm{FRI}i}=\boldsymbol{0}$ is set in the current paper.
$\boldsymbol{\tau}_{\mathrm{RBD,comp,}i}$ is the compensation of \ac{RBD} with $\boldsymbol{\tau}_{\mathrm{RBD,comp,}i}=\boldsymbol{\tau}_{\mathrm{RBD}i}$.

By combining (\ref{eq:imp_ctrl_law}) and (\ref{eq:dyn_short}) with $\boldsymbol{\tau}_{\mathrm{m}i}=-\boldsymbol{\tau}_{i}$, the following differential equation is obtained:
\begin{align}\label{eq:compensate_law}
	K_\mathrm{D}\dot{\boldsymbol{q}}_{\mathrm{m}i}=K_\mathrm{P}(\boldsymbol{q}_{\mathrm{cmd}i}-\boldsymbol{q}_{\mathrm{m}i})+J_i^T\boldsymbol{\lambda}_i
\end{align}

It is further investigated how the built-in controller of each individual robot $i=1,\dots,n_\mathrm{R}$ processes the joint position set-point provided by the user which is denoted as $\boldsymbol{q}_{\mathrm{sp}i}$.
It is discovered that $\boldsymbol{q}_{\mathrm{cmd}i}$ used in the control law (\ref{eq:imp_ctrl_law}) is not equal to the user input $\boldsymbol{q}_{\mathrm{sp}i}$, but results from a command filter, characterized as follows:
\begin{align}\label{eq:cmd_filter}
	q_{\mathrm{cmd}i}&=q_{\mathrm{sp}i}-K_{\mathrm{C}}\dot{q}_{\mathrm{sp}i}
\end{align}
The diagonal entries of the diagonal matrix $K_{\mathrm{C}}$ are identified in Table~\ref{tab:cmd_filter}.

\begin{table}[h]
	\caption{Identified parameters of the command filter}
	\begin{tabular}{lllllllll} 
			Joint & 1 & 2& 3&4&5&6&7\\ \hline
			Parameter in $10^{-3}\SI{}{\second}$ & 11.5 & 11.4 & 7.8&13.1&7.2&6.8&7.1
		\end{tabular}\label{tab:cmd_filter}
\end{table}

Given that $\boldsymbol{q}_{\mathrm{sp}i}$ is transmitted to the built-in controller from the user, it is essential for a DAE-MPC solver operating from the user side to generate $\boldsymbol{q}_{\mathrm{sp}i}$ as a result. 
Alternatively, when the solver generates $\dot{\boldsymbol{q}}_{\mathrm{sp}i}$, then $\boldsymbol{q}_{\mathrm{sp}i}$ can be transmitted to the built-in robot controllers immediately after an integration.
For the sake of building a complete \ac{DAE} model, the control input vector is selected to be $\boldsymbol{u}=[\dot{\boldsymbol{q}}_{\mathrm{sp}1}^T,\dots,\dot{\boldsymbol{q}}_{\mathrm{sp}n_\mathrm{R}}^T]^T$.
The differential variables are selected as $\boldsymbol{x}_\mathrm{d}=[\boldsymbol{q}_{\mathrm{m}1}^T,\dots,\boldsymbol{q}_{\mathrm{m}n_\mathrm{R}}^T,\boldsymbol{q}_{\mathrm{sp}1}^T,\dots,\boldsymbol{q}_{\mathrm{sp}n_\mathrm{R}}^T]^T$ because the derivatives of these variables occure in (\ref{eq:compensate_law}) and (\ref{eq:cmd_filter})

The differential equations can thus be formulated by combining (\ref{eq:compensate_law}) and (\ref{eq:cmd_filter}) in a matrix form:
\begin{align}\label{eq:diffeq}
	\dot{\boldsymbol{x}}_\mathrm{d}=A\boldsymbol{x}_\mathrm{d}+B\boldsymbol{u}+E\boldsymbol{x}_\mathrm{a}
\end{align}
The matrices in (\ref{eq:diffeq}) are as follows for the case of $n_\mathrm{R}=2$:
\begin{align}
	A=&\begin{bmatrix}-K_\mathrm{D}^{-1}K_\mathrm{P}&O&K_\mathrm{D}^{-1}K_\mathrm{P}&O\\
		O&-K_\mathrm{D}^{-1}K_\mathrm{P}&O&K_\mathrm{D}^{-1}K_\mathrm{P}\\O&O&O&O\\O&O&O&O\end{bmatrix}\\
	B=&\begin{bmatrix}-K_\mathrm{D}^{-1}K_\mathrm{C}&O\\
		O&-K_\mathrm{D}^{-1}K_\mathrm{C}\\I&O\\O&I\end{bmatrix}\\
	E=&\begin{bmatrix}O&O&K_\mathrm{D}^{-1}J_1^T&O\\
		O&O&O&K_\mathrm{D}^{-1}J_2^T\\O&O&O&O\\O&O&O&O\end{bmatrix}
\end{align}

The \ac{DAE} model of the physically coupled robots is comprised of the differential and algebraic equations (\ref{eq:diffeq}) (\ref{eq:alg}).
Thanks to the built-in joint specific impedance control that compensates \ac{RBD} of robots, the \ac{DAE} model is of index one without any derivatives of an order higher than one.

\section{Adaptive Model Predictive Control}
This section firstly handles the estimation of $\boldsymbol{p}$.
Afterwards, an \ac{MPC} based on the model with the updated $\boldsymbol{p}$ is formulated and solved to provide instant correction to the joint trajectory for an enhanced path accuracy.

\subsection{Estimation of the Uncertain Kinematic Parameters}\label{subsec:estim}
The basic idea of estimating $p$ is to compare the prediction of the DAE model with the sensor measurements of the real system and to update $p$ iteratively. The difference from the comparison is used in the following Newton-Raphson Root-Finding Method \cite{suli2003introduction}:
\begin{align}\label{eq:p_update}
	\boldsymbol{p}_{k+1}=\boldsymbol{p}_{k}-\alpha\left(\frac{\partial}{\partial\boldsymbol{p}}\boldsymbol{\tau}_\mathrm{m}\right)^{-1}\left(\tilde{\boldsymbol{\tau}}_\mathrm{m}-\boldsymbol{\tau}_\mathrm{m}\right)
\end{align}
Whereby, $k$ is the index of the iteration that updates $p$.
$0<\alpha\leq1$ is the step length of the update.
$\boldsymbol{\tau}_\mathrm{m}=[\boldsymbol{\tau}_\mathrm{m1}^T,\dots,\boldsymbol{\tau}_{\mathrm{m}n_\mathrm{R}}^T]^T$ is a concatenation of torque measurements of all robots, whereas $\tilde{\boldsymbol{\tau}}_\mathrm{m}$ is the prediction of the DAE model.
$\frac{\partial}{\partial\boldsymbol{p}}\boldsymbol{\tau}_\mathrm{m}$ is the sensitivity matrix of $\boldsymbol{\tau}_\mathrm{m}$ w.r.t. $p$ which is obtained from a sensitivity analysis \cite{andersson2018sensitivity}.
Among $\boldsymbol{x}_\mathrm{d}$ and $\boldsymbol{x}_\mathrm{a}$, only $\boldsymbol{\tau}_\mathrm{m}$, a part of $\boldsymbol{x}_\mathrm{a}$, is compared with its model prediction $\tilde{\boldsymbol{\tau}}_\mathrm{m}$ in (\ref{eq:p_update}) due to the following reasons.
In the DAE model, $\boldsymbol{p}$ only occurs in the algebraic equations (\ref{eq:alg}) instead of the differential equations (\ref{eq:diffeq}). 
W.r.t. changes of $\boldsymbol{p}$, the sensitivity of $\boldsymbol{x}_\mathrm{a}$ is much higher than that of $\boldsymbol{x}_\mathrm{d}$ because of the high magnitude of joint stiffness $K_\mathrm{J}$ that only appears inside $F$ in (\ref{eq:alg}) (\ref{eq:F_matrix}). 
The exact values of $K_\mathrm{J}$ are provided in \cite{busson2017task1}.
The comparison is further restricted to $\boldsymbol{\tau}_\mathrm{m}$ within $\boldsymbol{x}_\mathrm{a}$, because $\boldsymbol{\tau}_\mathrm{m}$ is measured directly by the torque sensors, which is not the case for $\boldsymbol{\lambda}$.

%Address the PE condition!

\subsection{Formulation of Model Predictive Control}\label{subsec:mpc}
The objective of the \ac{MPC} is to enhance the path tracking accuracy.
For a manufacturing process, the path is usually given in Cartesian space with up to six dimensions of translation and rotation. 
The path is input into the trajectory planning algorithm from the previous work \cite{ye2022stiffness}. 
It returns the optimal joint trajectories of all coupled robots without considering the uncertainties of kinematic parameters, $\boldsymbol{p}=\boldsymbol{0}$.
In the present paper, the generated joint trajectories are used as reference, denoted as $\boldsymbol{q}_\mathrm{ref}$. 
However, in the physically coupled configuration with uncertain kinematic parameters $\boldsymbol{p}\neq\boldsymbol{0}$, if robot $i$ tracks its own reference joint trajectory $\boldsymbol{q}_{\mathrm{ref}i}$, the other robots can no longer do the same.
Since robot $i=1$ is well calibrated and has its base on the origin, we assume that the error free forward transformation of its optimized reference trajectory $T_{\mathrm{kin}1}(\boldsymbol{q}_{\mathrm{ref}1})$ has been well planned in \cite{ye2022stiffness} and overlaps exactly with the manufacturing path, so the path accuracy problem is converted into the problem for robot $i=1$ to track $\boldsymbol{q}_{\mathrm{ref}1}$ with its link side joint positions. 
Therefore, we define the system output of the DAE as $\boldsymbol{y}=\boldsymbol{q}_{\mathrm{m}1}+\delta\boldsymbol{q}_{1}=\boldsymbol{q}_{\mathrm{m}1}+K_\mathrm{J}^{-1}\boldsymbol{\tau}_{\mathrm{m}1}$, in matrix form as follows:
\begin{align}\label{eq:output}
	\boldsymbol{y}=C\boldsymbol{x}_\mathrm{d}+G\boldsymbol{x}_\mathrm{a}
\end{align}
For two robots, the matrices can be shown in a tidy form:
\begin{align}
	C=&\begin{bmatrix}
		I&\boldsymbol{0}&\boldsymbol{0}&\boldsymbol{0}
	\end{bmatrix}\\
	G=&\begin{bmatrix}
	K_\mathrm{J}^{-1}&\boldsymbol{0}&\boldsymbol{0}&\boldsymbol{0}
\end{bmatrix}
\end{align}

It is worth noting that selecting $\boldsymbol{q}_{\mathrm{ref}1}$ as the only tracking target is not a master-slave control approach, because the control variables $\boldsymbol{u}$ are generated simultaneously for all robots, and the built-in controllers of robot $i=2,\dots,n_\mathrm{R}$ do not follow that of robot $i=1$.
In this way, the physically coupled system maintains the highest stiffness against $\boldsymbol{w}_\mathrm{ext}$ in manufacturing processes.

As such, the \ac{DAE}-\ac{MPC} with a receding horizon of $T$ beginning from $t_0$ can be formulated completely as follows:
\begin{subequations}\label{eq:mpc}
	\begin{flalign}
	\min_{\boldsymbol{u}}&\int_{t_0}^{t_0+T}\|\boldsymbol{u}-\boldsymbol{u}_\mathrm{ref}\|_Q^2+\|\boldsymbol{x}_\mathrm{a}\|_R^2+ \|\boldsymbol{y}-\boldsymbol{q}_{\mathrm{ref}1}\|_P^2\text{d}t\\
	&+V(\boldsymbol{y}_\mathrm{end}-\boldsymbol{q}_{\mathrm{ref}1\mathrm{,end}})\nonumber\\
	\text{s.t. }&\quad\text{DAE system dynamics (\ref{eq:diffeq}) (\ref{eq:alg}) (\ref{eq:output})}\nonumber\\
%	\text{s.t. }&\quad\dot{\boldsymbol{x}}_\mathrm{d}=A\boldsymbol{x}_\mathrm{d}+B\boldsymbol{u}+E\boldsymbol{x}_\mathrm{a}\\
%	&\quad F\boldsymbol{x}_\mathrm{a}=\boldsymbol{b}(\boldsymbol{p},\boldsymbol{w}_\mathrm{ext})\\
%	&\quad \boldsymbol{y}=C\boldsymbol{x}_\mathrm{d}+G\boldsymbol{x}_\mathrm{a}\\
	&\quad\dot{\boldsymbol{q}}_\mathrm{min}\leq\boldsymbol{u}\leq\dot{\boldsymbol{q}}_\mathrm{max}\label{eq:u_bound}\\
	&\quad\boldsymbol{q}_\mathrm{min}\leq\boldsymbol{x}_\mathrm{d}\leq\boldsymbol{q}_\mathrm{max}\label{eq:xd_bound}\\
	&\quad\boldsymbol{x}_\mathrm{d}(t_0)=\boldsymbol{x}_{\mathrm{d}0}\label{eq:initial}
\end{flalign}
\end{subequations}
Whereby, $Q, R$, and $P$ are weight matrices for the corresponding quadratic penalty terms, defined for any vector $\boldsymbol{z}$ and any weight matrix $L$ as $\|\boldsymbol{z}\|_L^2=\boldsymbol{z}^TL\boldsymbol{z}$. 
The trajectory error $\boldsymbol{y}-\boldsymbol{q}_{\mathrm{ref}1}$ is penalized with heavy weights because accuracy is the major objective of the controller.
$\boldsymbol{u}_\mathrm{ref}=\dot{\boldsymbol{q}}_{\mathrm{ref}}$ is obtained by a pre-processing of the reference trajectory $\boldsymbol{q}_\mathrm{ref}$.
The penalty on $\boldsymbol{u}-\boldsymbol{u}_\mathrm{ref}$ serves to let the optimizer provide a solution in the vicinity of the pre-processed reference  $\boldsymbol{u}_\mathrm{ref}$, instead of finding a distant one that requires an abrupt joint motion to reach. 
The penalty on $\boldsymbol{x}_\mathrm{a}$ optimizes the distribution of internal and external loads among the joints and over the coupler. 
$V(\boldsymbol{y}_\mathrm{end}-\boldsymbol{q}_{\mathrm{ref}1\mathrm{,end}})$ is the penalty function of the tracking error at the end of the horizon $t_0+T$.
$\boldsymbol{u}$ is bounded by joint velocity limits (\ref{eq:u_bound}), and $\boldsymbol{x}_\mathrm{d}$ is bounded by joint position limits (\ref{eq:xd_bound}).
Joint acceleration limits do not appear in the \ac{MPC}, but can be taken into consideration in the trajectory planning and the built-in controllers.
Finally in (\ref{eq:initial}), $\boldsymbol{x}_{\mathrm{d}0}$ stands for the initial state of $\boldsymbol{x}_{\mathrm{d}}$ that can be measured.

Problem (\ref{eq:mpc}) is solved with the direct collocation \cite{betts2010practical}.
\section{Results}
In this section, the parameter estimation in Section~\ref{subsec:estim} is tested on a pre-planned path with two physically coupled LBR robots in the real execution, $n_\mathrm{R}=2$.
The adaptive MPC in Section~\ref{subsec:mpc} is tested in the simulation based on the results from the real execution.

\subsection{Experimental Setting}
\begin{figure}[h]%[hbt]
%\begin{center}
	\begin{subfigure}{0.48\textwidth}
		\vspace*{2mm}
		\includegraphics[width=\textwidth]{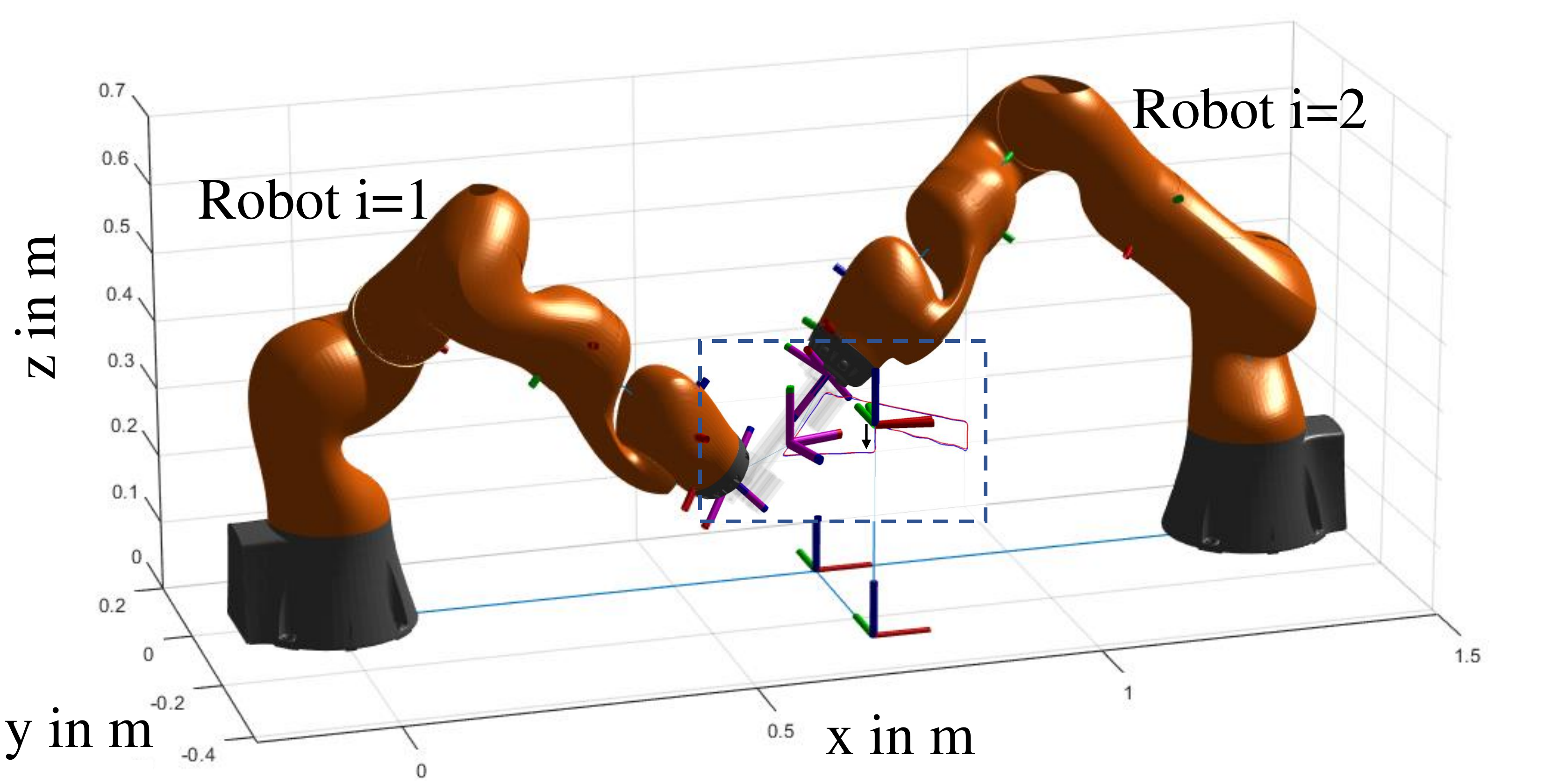}
		\caption{}%of{figure}{The loop closure condition of a two-robot example}
		\label{fig:ref_traj_rob}
	\end{subfigure}
%\end{center}
	\begin{subfigure}{0.49\textwidth}
		%		\vspace*{3mm}
		\include{fig/qref3d_new}
		\vspace*{-12mm}
		\caption{}
		%	\caption{Joint loads reduction by optimization}
		\label{fig:path_trans}
	\end{subfigure}
\caption{Subfigure (a) shows the overall scenario of the experiment both in simulation and real execution. Subfig.~(b) is the zoomed view of the enframed area in Subfig.~(a). The modified reference trajectories $\boldsymbol{q}_\mathrm{ref}$ for robot $i=1$ and $i=2$ are transformed into TCP positions in Cartesian space with the forward kinematics. The TCP positions start with the arrow, move clockwise, and finish a loop in $\SI{14.69}{\second}$}
\vspace*{-3mm}
\end{figure}

\begin{figure}[h]
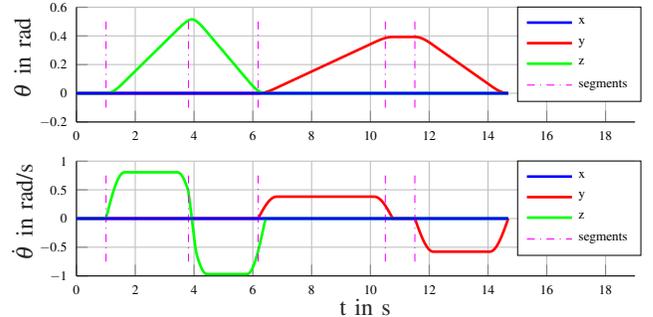
%[hbt]
	\begin{minipage}{.48\textwidth}
		%		\vspace*{3mm}
		\include{fig/path_angle}
		\vspace*{-9mm}
		\captionof{figure}{Reference Euler angles $\theta$ and angular velocities $\dot{\theta}$ of the coupler. The dash-dotted lines are the boundaries of segments where TCP comes to corners in Fig.~\ref{fig:path_trans}.}
		\vspace*{-3mm}
		%	\caption{Joint loads reduction by optimization}
		\label{fig:path_angle}
	\end{minipage}
\end{figure}

As shown in Fig.~\ref{fig:ref_traj_rob}, the two LRB robots are rigidly connected to the half-transparently visualized coupler. 
The base of robot $i=1$ is at the origin, while robot $i=2$ has the nominal base position $(\SI{1.365}{\meter},0,0)$.
In this preliminary study, the work-piece and the spinning spindle are not yet involved.
A nominal Cartesian path with six straight segments and five corners is placed between the two robots.
The corresponding reference joint trajectories are generated \cite{ye2022stiffness}. 
To investigate the response of the parameter estimation and the DAE-MPC to disturbances and vibrations both in simulation and real execution, the reference joint trajectories are superposed with a vibration in the second, fourth, and sixth segment.
So the resulting modified reference joint trajectories $\boldsymbol{q}_\mathrm{ref}$ with the vibration are no longer synchronized on the \ac{TCP}, as shown in Fig.~\ref{fig:path_trans}.
Meanwhile, $\boldsymbol{q}_\mathrm{ref}$ incorporates rotations of the coupler in the Cartesian space as shown in Fig.~\ref{fig:path_angle}. 
This is necessary because for the estimation of the rotational and translational kinematic parameters in $\boldsymbol{p}$, the condition of persistent excitation must be fulfilled \cite{erhart2016cooperative,narendra2012stable}.

In open-loop control, the built-in controllers of robots $i=1,2$ receive the corresponding $\boldsymbol{q}_{\mathrm{ref}i}$ without coordination. 
A cyclic internal stress will occure in the coupler due to the superposed vibration.
In the closed-loop control, the adaptive DAE-MPC should achieve the coordination between the built-in controllers by processing $\boldsymbol{q}_\mathrm{ref}$, outputing $\boldsymbol{u}=\dot{\boldsymbol{q}}_\mathrm{sp}$, and sending $\boldsymbol{q}_\mathrm{sp}$ to the built-in controllers.
%opengl hardware
%set(0,'DefaultLineLineSmoothing','on');
%set(0,'DefaultPatchLineSmoothing','on');
%opengl('OpenGLLineSmoothingBug',1);
%helps smoothing 3D plots

\subsection{Estimation of the Uncertain Kinematic Parameters}
As stated in Section~\ref{subsec:alg}, the uncertain kinematic parameters only include $\boldsymbol{p}=[\Delta\boldsymbol{p}_{\mathrm{b}2}^T,~^{\mathrm{body}}\Delta\boldsymbol{p}_{\mathrm{cp}2}^T]^T$ in case of two robots.
Considering the structure where the bases of the robots are mounted, the greatest uncertain kinematic parameters in $\Delta\boldsymbol{p}_{\mathrm{b}2}$ are the translational x- and z-components which can be over millimeters, whereas the other four components of $\Delta\boldsymbol{p}_{\mathrm{b}2}$ are at a lower order of magnitude. 
As of the uncertainty in the coupler geometry $~^{\mathrm{body}}\Delta\boldsymbol{p}_{\mathrm{cp}2}$, the rotational x- and z-components as well as the translational z-component are more critical.
Therefore, we reduce the dimension of estimated kinematic parameters to five and leave the rest seven components zero.

The estimation method is tested in a real execution with the open-loop control without MPC. 
The DAE system (\ref{eq:diffeq}) (\ref{eq:alg}) is solved by the integrator IDAS \cite{hindmarsh2005sundials}.
To compute the sensitivity matrix $\frac{\partial}{\partial\boldsymbol{p}}\boldsymbol{\tau}_\mathrm{m}$ in (\ref{eq:p_update}), IDAS provides the functionalities of sensitivity analysis within the framework of a CasADi plug-in \cite{frey2023fast}.
The update of (\ref{eq:p_update}) is conducted at a rate of $\SI{10}{\hertz}$ with a step length $\alpha =0.1$. 

\begin{figure}[h]
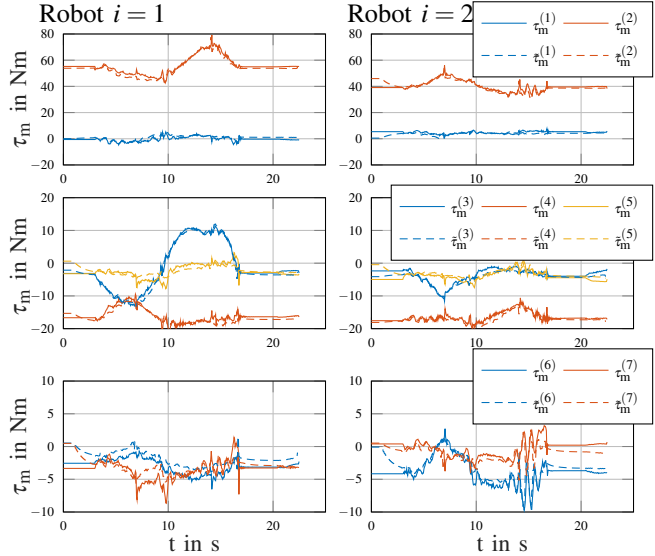
%[hbt]
	\begin{minipage}{.49\textwidth}
		%		\vspace*{3mm}
		\include{fig/joint_torque_sparse}
		\vspace*{-9mm}
		\captionof{figure}{Comparing the sensor measurement $\boldsymbol{\tau}_\mathrm{m}$ and model prediction $\tilde{\boldsymbol{\tau}}_\mathrm{m}$. They converge after starting the estimation of uncertain kinematic parameters from $t=\SI{1}{\second}$. Joint movements start at $t=\SI{3}{\second}$. In the legends, the joint numbers are given in superscript brackets.}
		%	\caption{Joint loads reduction by optimization}
		\label{fig:torque}
	\end{minipage}
\end{figure}

In Fig.~\ref{fig:torque}, it can be seen that before the error estimation is started at $t=\SI{1}{\second}$, there are significant differences between the torques from the measurement $\boldsymbol{\tau}_\mathrm{m}$ and the model prediction $\tilde{\boldsymbol{\tau}}_\mathrm{m}$ in both robots $i=1,2$.
After starting the estimation, the model prediction converges towards the measurement within $\SI{3}{\second}$.
This happens because the torque differences at the beginning are used to update the kinematic parameters of the DAE model, as shown in Fig.~\ref{fig:param_est}.

\begin{figure}[h]%[hbt]
	\begin{minipage}{.49\textwidth}
				\vspace*{-5mm}
		\include{fig/param_est}
		\vspace*{-9mm}
		\captionof{figure}{Estimation of rotational and translational components of $\Delta\boldsymbol{p}_{\mathrm{b}2}$ and $~^\mathrm{body}\Delta\boldsymbol{p}_{\mathrm{cp}2}$, starting from $t=\SI{1}{\second}$}
		%	\caption{Joint loads reduction by optimization}
		\label{fig:param_est}
	\end{minipage}
\vspace*{-5mm}
\end{figure}

From Fig.~\ref{fig:torque} it can be further observed that the joint torques from the measurement and the model prediction exhibit the same trend and overlap mostly.
The proposed DAE model (\ref{eq:diffeq}) (\ref{eq:alg}) simulates the load behavior of the real multi-robot system at a decent quality after an effective parameter estimation. 
However, the discrepancy in torques does exist as the measurement skips up and down over the model prediction.
This phenomenon occurs with a consistent pattern, only when the corresponding joint velocity changes direction.
Therefore, the skipping of the torque measurement is attributed to the Coulomb friction in the joints that is neglected in the model.
Further causes of the discrepancy include the neglected elasticity of the coupler and the robotic links.
All the above simplifications in the model limit its prediction ability to the quasi-statics region of the system dynamics.
This leads to the fluctuation of the estimated parameters in Fig.~\ref{fig:param_est}.
For a more powerful parameter estimation, one can either reduce the step length $\alpha$ or perform a dynamic identification of the robots so as to obtain more accurate RBD to be used in vector $\boldsymbol{b}(\boldsymbol{p},\boldsymbol{w}_\mathrm{ext})$ of (\ref{eq:alg}).

\subsection{Adaptive Model Predictive Control}
\begin{figure}[h]
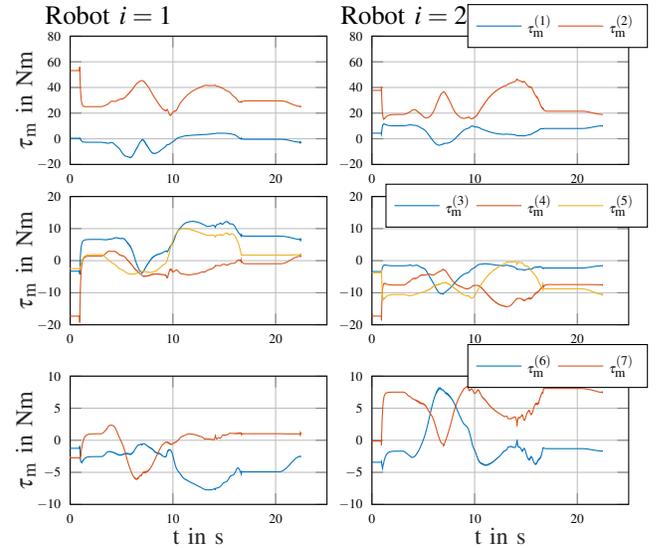
%[hbt]
	\begin{minipage}{.48\textwidth}
		%		\vspace*{3mm}
		\include{fig/controlled_torque_format}
		\vspace*{-9mm}
		\captionof{figure}{Joint torques with the adaptive MPC started at $t=\SI{1}{\second}$ and the joint movements started at $t=\SI{3}{\second}$.}
		%	\caption{Joint loads reduction by optimization}
		\label{fig:torque_mpc}
	\end{minipage}
\end{figure}

With the parameter estimation in place, the adaptive MPC is simulated on $\boldsymbol{q}_\mathrm{ref}$ that has been superposed with vibrations as in Fig.~\ref{fig:path_trans}.
The resulting joint torques are shown in Fig.~\ref{fig:torque_mpc}.
Before starting the adaptive MPC at $t=\SI{1}{\second}$, the joint torques are close to the measured values in Fig.~\ref{fig:torque}.
But afterwards, a more even distribution is achieved.
A significant load reduction can be observed among the joints that originally bear high torques.
More torques are shifted to joint~1, 6, and 7 which originally bear smaller loads. 
Through a further comparison between Fig.~\ref{fig:torque} and Fig.~\ref{fig:torque_mpc}, the adaptive MPC has a damping effect on the cyclic loads that are originally induced by the superposed vibration of $\boldsymbol{q}_\mathrm{ref}$.

\begin{figure}[h]
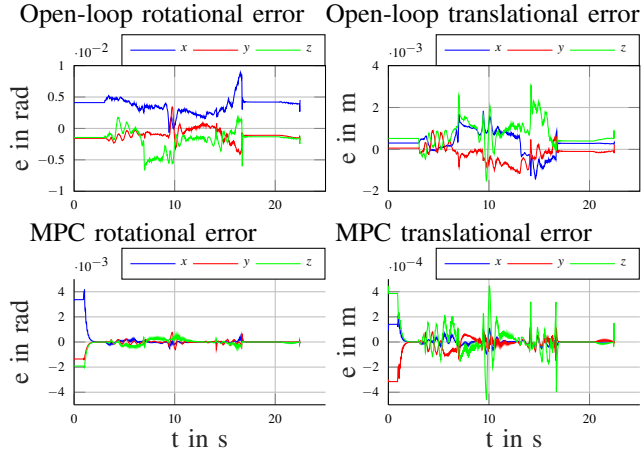
%[hbt]
	\begin{minipage}{.47\textwidth}
		%		\vspace*{3mm}
		\include{fig/control_err}
		\vspace*{-9mm}
		\captionof{figure}{Rotational and translational errors of TCP under open-loop control and closed-loop adaptive MPC.}
		%	\caption{Joint loads reduction by optimization}
		\label{fig:error_mpc}
	\end{minipage}
\end{figure}

The multi-robot system is able to track the still fluctuating $\boldsymbol{q}_{\mathrm{ref}1}$ with a high accuracy, as shown in Fig.~\ref{fig:error_mpc}.
As the superposed vibration is at the highest amplitude of $\SI{2.4}{\milli\meter}$ at $t=\SI{9.75}{\second}$, shown at the beginning of the fourth segment in Fig.~\ref{fig:path_trans}, the greatest error of the adaptive MPC appears.
Its translational component is $\SI{0.463}{\milli\meter}$ which is reduced by 84.8\% compared to the open-loop control, and the rotational component $8.93\mathrm{e}{-3}\SI{}{\radian}$ is reduced by 91.6\%.
Over the whole trajectory, the average translational error is $\SI{0.0663}{\milli\meter}$, and rotational $0.159\mathrm{e}{-3}\SI{}{\radian}$, with a reduction of 93.5\% and 96.5\% respectively.
The only state-of-the-art benchmark of a comparable scenario is the previous work \cite{ye2023enhancement}, compared to which the average translational error is reduced by 88.6\%.

Although the adaptive MPC shows high performance in the simulation, an implementation on the real robots is not yet accomplished due to the real-time capability issue of the collocation method on the processor Intel Core i7-10700K CPU@3.80GHz$\times$16.
For a horizon of $T=\SI{0.1}{\second}$ and a step interval of $\SI{0.01}{\second}$, the collocation is solved in $\SI{0.404}{\second}$ in Matlab and $\SI{0.294}{\second}$ in C++.

\section{Conclusion and Outlook}
The present work proposes an adaptive DAE-MPC method to achieve a high path accuracy online. 
The DAE model incorporates the uncertain kinematic parameters, the limited stiffness of robots, and the built-in robot controller that cannot be circumvented. 
The uncertain kinematic parameters are estimated effectively with measurement data from real execution, resulting in a successful model prediction.
The MPC achieves an even load distribution and a high tracking accuracy on disturbed reference trajectories in simulation.

In the future work, a more efficient solution method should be applied to the adaptive DAE-MPC, so as to facilitate an online execution on the real robots. 
The torque overlay of the built-in robot controller can be exploited for a higher stiffness in manufacturing processes.

\addtolength{\textheight}{-12cm}   % This command serves to balance the column lengths

%%%%%%%%%%%%%%%%%%%%%%%%%%%%%%%%%%%%%%%%%%%%%%%%%%%%%%%%%%%%%%%%%%%%%%%%%%%%%%%%

%%%%%%%%%%%%%%%%%%%%%%%%%%%%%%%%%%%%%%%%%%%%%%%%%%%%%%%%%%%%%%%%%%%%%%%%%%%%%%%%

%%%%%%%%%%%%%%%%%%%%%%%%%%%%%%%%%%%%%%%%%%%%%%%%%%%%%%%%%%%%%%%%%%%%%%%%%%%%%%%%
%\section*{APPENDIX}

\section*{ACKNOWLEDGMENT}
The authors thank Stefan Schwab for insightful discussions, Tobias Sch{\"u}rmann and Max Grobbel for valuable comments and suggestions to improve the manuscript.
%Avoid the stilted expression, One of us (R. B. G.) thanks . . .  Instead, try R. B. G. thanks. Put sponsor acknowledgments in the unnumbered footnote on the first page.

\renewcommand*{\bibfont}{\footnotesize}
\printbibliography

\end{document}